\documentclass[aps,prb,reprint,preprintnumbers]{revtex4-1}
\usepackage{graphicx}
\usepackage{bm}
\usepackage{epsfig}
\usepackage{ulem}
\usepackage{color}
\usepackage{mathrsfs}
\usepackage{dcolumn}
\usepackage{setspace}
\usepackage{array}
\usepackage{amsmath}
\usepackage{amssymb}
\usepackage{dsfont}
\usepackage{gensymb}
\usepackage{dcolumn}
\usepackage{multirow}
\usepackage{bibentry,natbib}
\usepackage{booktabs}
\usepackage{color}

\newcommand {\be}[1]{
      \begin{eqnarray} \mbox{$\label{#1}$}  }

\newcommand{\ee}{\end{eqnarray}}

\begin{document}
\bibliographystyle{unsrt}
\title{Geometric Induction in Chiral Superconductors}

\author{Qing-Dong Jiang$^1$, T. H. Hansson$^1$, Frank Wilczek${^1}{^2}{^3}{^4}$}
\affiliation{{}\\ $^1$Department of Physics, Stockholm University, Stockholm SE-106 91 Sweden\\
$^2$Center for Theoretical Physics, Massachusetts Institute of Technology, Cambridge, Massachusetts 02139 USA\\
$^3$Wilczek Quantum Center, Department of Physics and Astronomy, Shanghai Jiao Tong University, Shanghai 200240, China\\
$^4$Department of Physics, Arizona State University, Tempe AZ 25287 USA}
\begin{abstract}
We consider a number of effects due to the interplay of superconductivity, electromagnetism and elasticity, which are unique for thin membranes of  layered chiral superconductors. Some of them should be within the reach of present technology, and could be useful for characterizing materials. More speculatively, the enriched control of Josephson junctions they afford might find useful applications.
\end{abstract}
\preprint{MIT-CTP/5166}
\maketitle

\textit{Introduction}:
In spite of its long history, superconductivity is still a very active area of research with promises both of new  applications and profound physics insights. The origin of this open-ended story is the immense variety of pairing mechanisms and pairing channels. For example, the high $T_c$ cuprates feature $d$-wave pairing whose mechanism is at present unclear. 

Recently there has been great interest in a class of superconductors called ``chiral'' superconductors, whose order parameter, involving angular momentum, breaks both time reversal and parity.  Several promising candidates have been proposed, including $\rm Sr_2RuO_4$ \cite{maeno} (chiral $p$-wave), $\rm SrPtAs$\cite{sigrist}, doped or twisted-bilayer graphene\cite{kiesel,ccliu} (chiral $d$-wave), and $\rm UPt_3$\cite{htou} (chiral $f$-wave).  
The special interest in the odd chiral pairing materials is largely due to the prediction that they support Majorana zero modes in vortex cores and at edges\cite{kopnin,volovik}, which are central to several proposals for topological quantum computation \cite{nayak}.  However, despite intensive theoretical study and  huge experimental efforts, no material has been proven definitively to be a chiral superconductor \cite{mackenzie}. In this Letter, we study various phenomena in thin films of chiral superconductors that depend on the geometry of the sample. The physical realization can either be very thin films, or strongly layered materials, where the physics is essentially two dimensional. The crucial assumption is that the motion of the electrons in a Cooper pair is in the plane and that the curvature of the sample is small enough for the Cooper pair to be approximated as a point particle with spin perpendicular to the surface. The geometric phases that arise when these spins are transported along paths on a curved surface will be at the heart of the phenomena we shall study.  They define unambiguous, and seemingly accessible, signatures for chiral superconductivity.   Closely related time-dependent effects promise to afford additional control of Josephson junctions through mechanical coupling.  

Effects of the background geometry on paired 2D phases of matter have been studied earlier, both for thin films of Helium III\cite{vollhardt} and for $p$-wave superconductors\cite{kvorning,hoyos,spanslatt}. Those studies were limited to static backgrounds.  In this Letter, we shall extend the analysis to the temporal regime. 
Perhaps surprisingly, this not only allows us to study the electromagnetic response of a time-dependent (spatial) geometry but also reveals a more subtle {\it ac geometric Josephson effect} that occurs when rotating a flat geometry. Although the energy scales characteristic of the superconducting condensate is typically much smaller than the elastic energies one can envision situations, involving very thin films, where that is not the case. 
To analyze such situations we derive a theory for a thin film of a layered chiral 2D superconductor where both the response of the chiral superconducting condensate to changes in the geometry (i.e., strain) and the elastic response of the material are taken into account.  

The Letter is organized as follows: We first develop the necessary formalism, recall some earlier results for fixed geometries, and propose an embodiment of the geometric Josephson effect that appears more practical than previous suggestions. 
Turning to steady-state rotations, we will introduce the rotational ac Josephson effect.   Finally we turn to  chiral superconductivity on flexible geometries, where the interplay between geometric deformation and magnetism is exemplified in several phenomena and devices.  

\textit{Superconductivity, electromagnetism, and elasticity}: The order parameter of a chiral $p$-wave superconductor is a complex vector defined in the tangent plane of a surface. It can be written as
$
\Psi=\psi~\epsilon_\pm,
$
where $\epsilon_{\pm}=\left(\hat e_1\pm i\hat e_2\right)/\sqrt{2} $ are chiral basis vectors defined in terms of the local orthogonal basis vectors, or zweibeins, $\{\hat e^a \}$, $a=1,2$ and $\hat e_a = \hat e^a$. The $\pm$ sign denotes the chirality, and  $\psi=\sqrt{\rho} e^{i\theta}$ is the complex amplitude in terms of  the superfluid density $\rho$ and phase $\theta$. Generalization to a chiral $l$-wave order parameter, describing a condensate of Cooper pairs with orbital angular momentum $l\hbar$ is straightforward\cite{volovik1}.

On a curved surface, the simplest effective model of a chiral $l$-wave superconductor is that of a time-dependent Ginzburg-Landau (GL) action properly coupled to electromagnetism and the background geometry,
\begin{equation}
\mathcal L_{sc}=i\hbar\psi^*D_t\psi-\frac{\hbar^2 g^{ij}}{2m}(D_i \psi)^*(D_j \psi)-V(|\psi|) \, ,
\end{equation}
where $V(|\psi|)$ is a potential causing spontaneous symmetry breaking in the chirality + channel and the covariant derivative is defined by
\begin{equation}
D_\mu=\partial_\mu-i\frac{q}{\hbar} A_\mu+il\, \omega_\mu,
\end{equation}
with the electromagnetic potential  $A_\mu=(-\phi, A_x, A_y)$, $q = 2e$ as the Cooper pair charge, and $\omega_\mu=\hat e_1\cdot \partial_\mu\hat e_2$ as the spin connection. Also, $g^{ij}$ is  the inverse of the metric tensor $g_{ij} =e^a_i e_{a,j} $.  

To describe a flexible thin membrane, we adopt the Monge representation, $R(x,y)=[x,y ,h(x,y)]$, to parametrize a two dimensional surface embedded in three-dimensional space, and take the free energy functional 
\begin{eqnarray}\label{fgeo}
 H_{geo}=\int d^2 r\left(\frac{\kappa}{2}  \mathscr H^2-\kappa_G K\right) \, ,
\end{eqnarray}
to describe the elastic response\cite{nelson}. Here $\mathscr H$ is the extrinsic curvature and  
\begin{equation}
K  ~=~ \frac{1}{2} \frac {\epsilon^{ij}} {\sqrt{g} } \,  (\partial_i \omega_j - \partial_j \omega_i  ) ~\equiv~  \frac{1}{2} \frac{\epsilon^{ij}}  {\sqrt{g} }R_{ij}
\end{equation}
is the Gaussian curvature of the surface; $R_{ij}$ is the Riemann curvature,  and $g = | \mathrm {det}\, g_{ij}|$.
$\kappa$ and $\kappa_G$ are the bending and Gaussian rigidity, respectively. A rigid planar geometry corresponds to $\kappa, \kappa_G \rightarrow \infty$. 
In the small gradient approximation ($|\nabla h|<<1$), the extrinsic curvature $\mathscr H\approx \nabla^2 h$, the Gaussian curvature $K\approx \partial_x^2 h\partial_y^2 h-(\partial_x\partial_y h)^2$, and the spin connection $\omega_i \approx \frac{1}{2}\epsilon^{kl}\partial_l[(\partial_k h)(\partial_i h)]$ \cite{nelson2}.
Note that since the Gaussian curvature term $\sqrt{g}K$ is a total derivative it will not affect the local equations of motion of the system, but it comes into play for nontrivial topologies. 

For most of what follows it will be sufficient to use the approximate Hamiltonian  $ H = \int d^2 r\, \sqrt{g}\,  \mathcal H$ defined as
\begin{eqnarray} \label{totham}
\mathcal H&=&\mathcal H_{sc}+\mathcal H_{geo}+\mathcal H_{em}\nonumber\\
&=& \gamma_0  \left(\partial_0\theta-\frac{q}{\hbar} A_0+l \, \omega_0 \right)^2  + \frac{\gamma}{2}  \left(\vec\nabla\theta-\frac{q}{\hbar}\vec A+l \, \vec\omega \right)^2  \nonumber\\
&&+\frac{\kappa}{2 \sqrt g} \left(\nabla^2 h\right)^2+\frac{d_0}{2\mu_0}  B^2.
\end{eqnarray}
where $\gamma=\frac{\hbar^2 \rho}{m}$ is the stiffness of the superconductor,  $B=\frac{\epsilon^{ij}}{\sqrt{g}}\partial_i A_j$  is the magnetic field pseudoscalar, $d_0$  is the thickness of the sample, and $\mu_0$ is the magnetic permeability. The term $\sim\gamma_0$ was obtained by integrating out the fluctuations of the superconducting density. $\mathcal H$ does not include any kinetic energy for the membrane or the electromagnetic fields, nor electric field energy.  Thus the geometry is treated quasistatically, and we assume that charge fluctuations and electric fields are small.   With the density fixed, only the phase field $\theta$ remains to control the low-energy dynamics of the superconducting condensate; of course, that assumption fails near vortex cores, around which $\theta$ exhibits winding singularities.
Use of $B^2$ for the magnetic energy is appropriate for a film of a layered material  whose thickness $d_0$ is much larger than the London length $\lambda_L=\sqrt{\frac{\hbar^2 d_0}{q^2 \mu_0\gamma}}$.  Later we will consider some situations where that condition does not hold, and then we will discuss the necessary modifications of the above description.

The Hamiltonian \eqref{totham} captures both the  Hall viscosity\cite{hoyos} and the  geometric Meissner effect\cite{kvorning} of a chiral superconductor.  Let us briefly recall the latter.

From \eqref{totham} we find the supercurrent density,
\begin{equation}
J_i = - \frac{\delta \mathcal L_{sc}}{\delta A^i}=\frac{\rho \hbar}{m}\left(\partial_i\theta-\frac{q}{\hbar} A_i+ l \omega_i\right).
\end{equation}
For a fixed geometry the energy  $\mathcal H_{sc}$ is minimized when the current vanishes.  Taking the curl, we find
\begin{equation}\label{curvatures}
qB = l\hbar \sqrt g K.
\end{equation}
This relation shows that a magnetic field can be induced from a Gaussian curvature. Formally, we also have delta-function contributions at vortex cores.  An integrated form of \eqref{curvatures} can be derived with fewer approximations, because we can use Stokes' theorem assuming only that $J$ vanishes along the boundary.  This is the geometric Meissner effect described at length in Ref. \onlinecite{kvorning}.

\textit{Geometric dc Josephson effect}: We can exploit the tight relation between magnetic field and geometric curvature to derive a diagnostic for chiral superconductivity.  

A Josephson junction (JJ) penetrated by a magnetic flux displays a supercurrent diffraction pattern \cite{tinkham}:  
\be{jjrel}
I_c(B)=I_c(0)  \frac{\sin(\pi \Phi_B/\Phi_0)}{\pi \Phi_B/\Phi_0}.
\ee
where, $\Phi_B$ is the  magnetic flux,  $\Phi_0 $  the flux quantum and $I_c(B)$ the maximum net tunneling current.

\begin{figure}[!htb]
\includegraphics[height=3cm, width=8cm, angle=0]{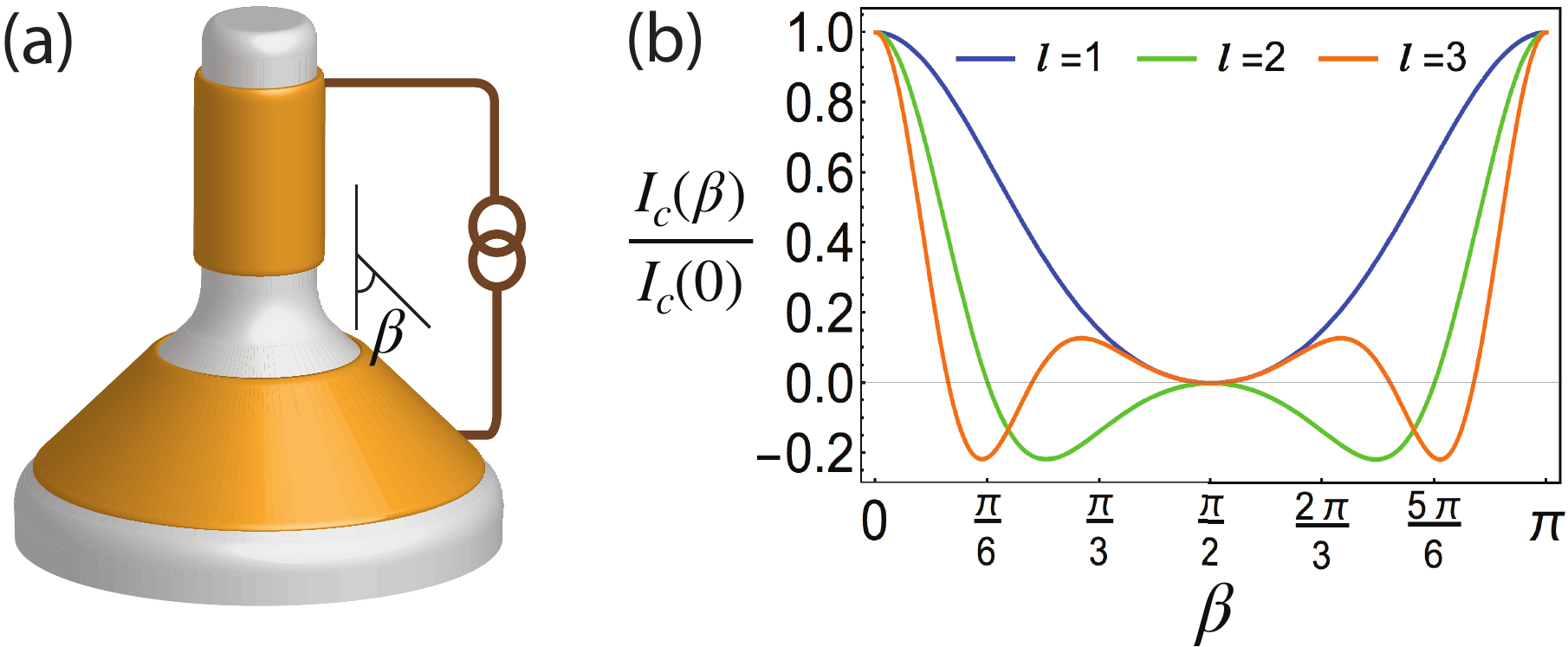}
\caption{(a) Schematic structure of the geometric Josephson junction. The yellow layers represent chiral superconducting electrodes on top of a metallic substrate (gray). $\beta$ represents the tilting angle of the junction. (b) Tunneling supercurrent is a function of tilting angle $\beta$.  \label{fig1}}
\end{figure}

An analogous effect occurs if the junction supports an integrated curvature, as displayed in Fig. \ref{fig1}(a). The spin connection one-form on the cone is $\omega = \sin\beta\, d\varphi$, where  
$\varphi$ is the  azimuthal angle, while it vanishes on the cylinder\cite{quelle}. The total curvature of the junction is $K_J = 2\pi\, \sin\beta$, and thus in analogy with \eqref{jjrel} we find
\begin{equation}
I_c(\beta)=I_c(0)\frac{\sin(l\pi\sin\beta)}{l\pi\sin\beta}.
\end{equation}
Depending on the pairing channel $l$, the tunneling current vanishes at different critical angles. For a chiral $p$-wave superconductor ($l=1$) the tunneling current vanishes at the zeroing angle $\beta=\pi/2$, regardless the phase difference across the JJ. For chiral $d$-wave pairing the zeroing angle is $\beta=\pi/4$ etc.  
The vanishing of the Josephson tunneling current at a specific zeroing angle is a clear signature for chiral superconductivity and allows for unambiguous determination
of the pairing channel.  The critical tunneling current as a function of  the tilting angle is shown in Fig. \ref{fig1}(b).
Note that our proposal is essentially different from previous studies considering flat JJs of chiral superconductors \cite{barash}. 
 
\begin{figure}[!htb] 
\includegraphics[height=3.0cm, width=2.6cm, angle=0]{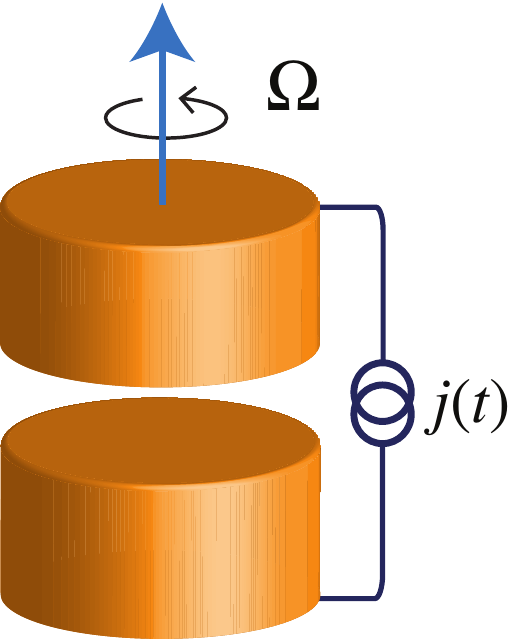}
\caption{Schematic structure of rotational ac Josephson junction.  \label{fig2} }
\end{figure}

\textit{Rotational ac Josephson effect}: When a JJ is  biased at a constant voltage, $A_0$, the current will vary periodically with the period $\Omega_0 = 2eA_0/\hbar$; this is the ac Josephson effect. Analogously there is a geometric ac Josephson effect if one element of a junction is rotated at a constant speed relative to the other, as displayed in Fig. \ref{fig2}. In the laboratory frame the zweibeins rotate as $\hat e_1=\cos \Omega t ~\hat e_x-\sin \Omega t ~\hat e_y$ and $\hat e_2=\sin\Omega t ~\hat e_x+\cos \Omega t ~\hat e_y$, resulting in a time component of the spin connection, $\omega_0=\hat e_1\cdot \partial_t\hat e_2= \Omega$. From \eqref{totham} we infer that the effective voltage drop - i.e., the chemical potential difference -  across the JJ  is the sum of the electric potential part $V$ and geometric potential part $l\hbar \Omega/2e$.  (The ``geometric'' part can also be derived in a different and very general way, by noting that the angular momentum operator is the generator of rotations.)   Thus, relative rotation is sufficient to drive an ac Josephson tunneling current at frequency $l\Omega$, even in the absence of an applied voltage.   This provides another method to detect chirality and pin down the pairing channel. 
From an alternative viewpoint, the rotating JJ works as an ideal {\it electricity generator} that converts mechanical energy to electrical power without any internal ohmic dissipation.


The usual ac Josephson effect can be measured either directly from the emitted microwave radiation\cite{radiation}, or from the 
Shapiro steps in the I-V characteristics when applying radiation\cite{shapiro}, but it is difficult to directly measure the varying tunneling current
since a small voltage corresponds to a high-frequency oscillation ($\sim 500$ GHz/mV). By sharp contrast, the rotational ac Josephson effect, once realized, gives rise to a tunneling current at much lower frequencies, under mechanical control.  Of course, one can also combine electric voltage and mechanical rotation, allowing fine modulation of the former. 
Rotation will also give rise to a shift in the Shapiro steps, but  even a rotation frequency of 1000 Hz, amounts to a voltage shift in the pV range, which is at the very limit of current metrology.


By rotating a superconductor in liquid Helium one has measured the London moment\cite{hilde}.
The corresponding static magnetic field  will not affect the ac frequency, but can in principle modify the tunneling current. 
For a JJ with an area $\sim (\mu m)^2$, the corresponding flux is only a small fraction of a flux quantum, so the effect on the tunneling current will not be significant. It would be technically challenging to realize rotating JJs, but their remarkable properties, for chiral superconductors, offer significant motivation.

\textit{Membrane quasistatics}: We now consider a flexible thin membrane, where geometric curvature and magnetic field can be mutually induced.  Varying the total Hamiltonian {$\mathcal H$} in \eqref{totham} with respective to $\theta$, $\vec A$ and $h$ give the GL equations:
\begin{eqnarray}\label{gl1}
 \frac{1}{\sqrt{g}}\partial_i(\sqrt{g} J^i) &=&0 \ \ \ ;  \ \ \ 
 \frac{d_0}{\mu_0}(\epsilon^{ij}\partial_j B) =\sqrt{g}J^i  \\
 \kappa\nabla^4 h &=&\hbar\frac{l}{q} (\partial_i \partial_j h) \,\epsilon^{jk}\partial_k (\sqrt{g} J^i) \, ,
\end{eqnarray}
where the supercurrent density is
\begin{equation}\label{glcurrent}
J^i=\frac{q\gamma}{\hbar} g^{ij}\left(\partial_j\theta-\frac{q}{\hbar} A_j+l\, \omega_j\right) \, .
\end{equation}
Equations \eqref{gl1}  are not independent, since the second  implies the first one, and the 
above GL equations can  be reformulated as 
\begin{eqnarray}\label{gl3ac}
\nabla^2 B &=&-\frac{\hbar}{q \lambda_L^2} \left(l K-\frac{q}{\hbar} B\right)\\
\nabla^4 h &=&  \frac{l q \lambda_G^2}{\hbar} \left[\partial_i^2 h\partial_j^2 B-(\partial_i\partial_j h)(\partial_i\partial_j B)\right],
\end{eqnarray}
where we introduced the  elastic bending length $\lambda_G=\sqrt{\frac{\hbar^2 d_0}{q^2 \mu_0 \kappa}}$, which is, in addition to the London length $\lambda_L$, a characteristic length scale in the system. 
For nonchiral superconductors ($l=0$), the two equations decouple, yielding separate equations for the elastic and electromagnetic response. By contrast, for chiral superconductors, the two equations are coupled, resulting in analytically intractable fourth-order coupled differential equations. 

As an example, let us consider the elastic and magnetic response of thin circular  superconducting membrane that is clamped at its circumference in such a way that the tilt angle $\alpha = h'(R)$ (where $R$ is the radius of the circle) is kept fixed. In the absence of any magnetic field, the shape of the membrane is obtained by minimizing the elastic energy keeping $\alpha$ fixed.  In particular, it will be flat when $\alpha = 0$.   If a magnetic field is applied along the symmetry axis, the field will penetrate the distance $\lambda_L$ from the circumference. When the tilt angle $\alpha$ is non-zero, the membrane will be curved, and if the superconductor is chiral there will be a nonzero magnetic field deep inside the membrane due to the geometrical Meissner effect. Both the shape of the membrane, and the profile of the magnetic field will differ from the $\alpha = 0$ case. 

\begin{figure}[!htb]
\includegraphics[height=3.cm, width=8cm, angle=0]{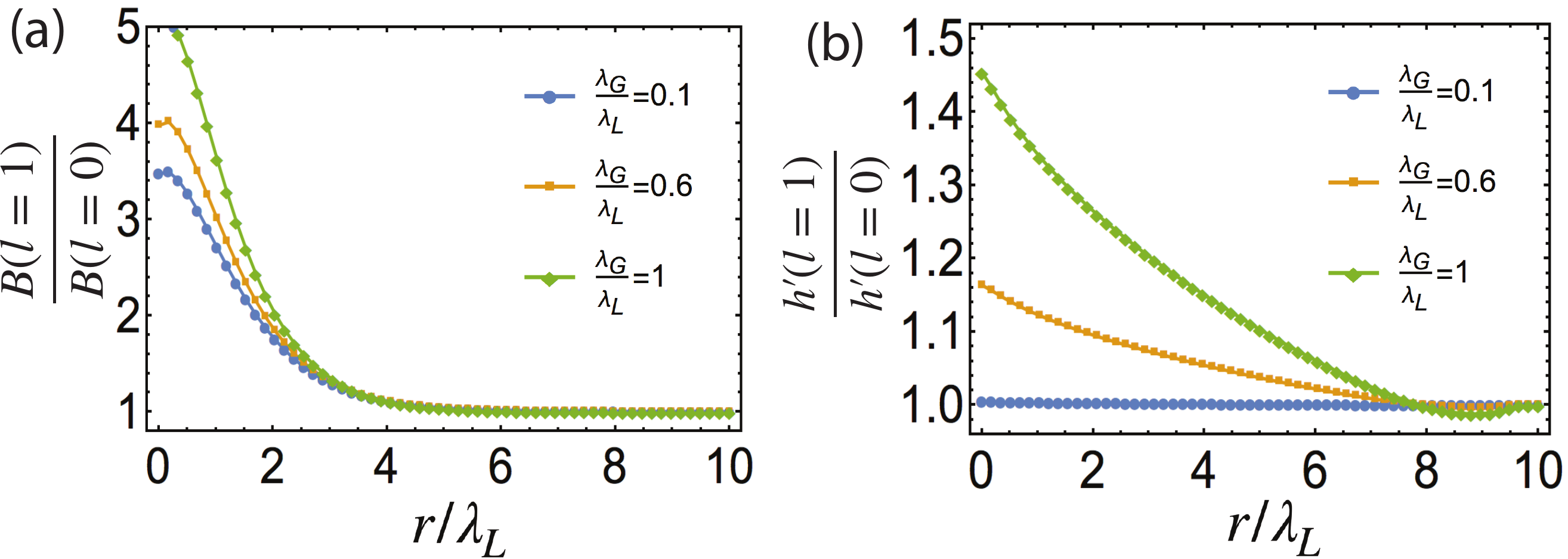}
\caption{Numerical simulation of spatial profiles of the magnetic field (a) and the height gradient (b). Different colors represent different bending lengths $\lambda_G$. The boundary conditions are set as $B(r=10 \lambda_L)=\hbar \lambda_L/q$ and $\alpha = h^\prime(r=10\lambda_L)=-0.2$. \label{fig3} }
\end{figure}

To quantify this \textit{chiral magnetostriction effect}, we have numerically solved the GL equations, for various ratios of the characteristic lengths $\lambda_L$ and $\lambda_G$. Fig. \ref{fig3}(a) compares the magnetic field distribution for a $l=1$  chiral superconductor to that for a conventional superconductor. 
The softer (larger $\lambda_G$) the material is, the more deeply the magnetic field can penetrate, since mechanical deformation provides an additional way to relieve phase frustration due to the magnetic field.  Fig.  \ref{fig3}(b) shows how the slope of the height  deviates from the  case where the magnetic field and the geometry are decoupled. 
The parameter range in Fig. \ref{fig3} is not entirely academic; for example, the geometric bending length is $\lambda_G=1\, \mu$m for stack-layered graphene with bending rigidity $\kappa\approx 1$ eV and thickness $0.1 \, \mu$m \cite{oppen}, and such  a $\lambda_G=1 \, \mu$m  is comparable with the London  length.

\bigskip

Now we describe two devices that utilize the interplay between chiral superconductivity and elasticity in an ultrathin film, where $\lambda_L \gg d_0$. 
In this case the magnetic energy dominates the energy of current flow, so there will be no geometrical Meissner effect. Instead, the curvature is unscreened.

\begin{figure}[!htb]
\includegraphics[height=3.2cm, width=7cm, angle=0]{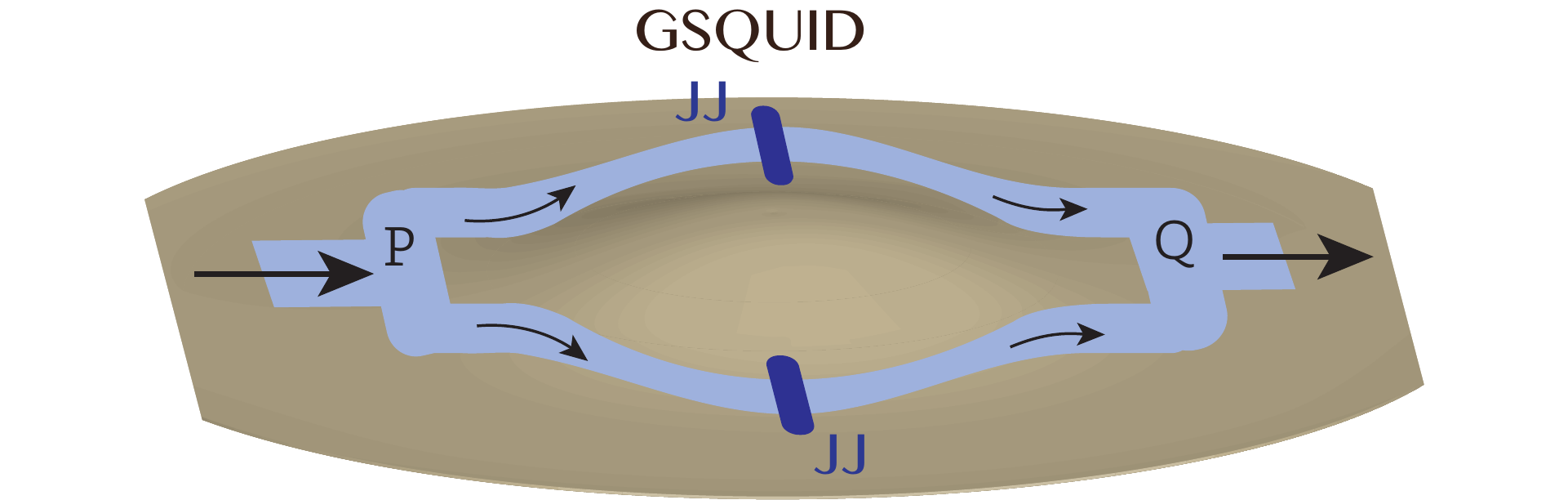}
\caption{A schematic picture of the  GSQUID. A finite curvature is enclosed by microribbons of a chiral superconductor (light blue), with two Josephson junctions (deep blue). Black arrows show the  direction of the supercurrent. \label{fig4}}
\end{figure}

\textit{Geometric SQUID}:

This device, displayed in  Fig. \ref{fig4}, consists of  
a very thin chiral superconducting film shaped into a SQUID configuration and  deposited on a flexible insulating membrane. It measures the parallel transport phases due to geometric deformation directly. We will refer to it as a geometric SQUID, or GSQUID.

Assuming the superconducting strips are narrow compared to the radii of curvature, the  phase difference between the points  P and Q in the picture can  be calculated either through the upper path or the lower path:
\be{geomsq}
\Delta\theta=\delta_a+l \int_{\Gamma_1}  \omega_i d l^i  =\delta_b+l \int_{\Gamma_2}  \omega_i d l^i\, , 
\ee
where $\delta_a$ and $\delta_b$ are the phase jump across the JJs. Thus we get
\begin{eqnarray}
\delta_b-\delta_a=-l\oint_\Gamma    \omega_i  d l^i,
\end{eqnarray}
where $\Gamma = \Gamma_1-\Gamma_2$ is a clockwise closed loop inside the superconducting ribbon. Next, define $2\delta_0 = \delta_a + \delta_b$, to get
\be{mechSQ}
J_t =J_0 (\sin \delta_a+\sin \delta_b)
=2J_0 \sin\delta_0 \cos \left(\frac{l}{2}\Phi_G\right )
\ee
for the total current through the junction. Here $\delta_0$ depends on the details of the junctions, and $\Phi_G$ is the total Gaussian curvature of the area of the substrate enclosed by the ribbon. Just as a normal SQUID is used to detect small changes in the magnetic field, the GSQUID can detect small changes in the  curvature of the part of the membrane inside the GSQUID loop. Such geometric changes can be induced  by acoustic driving or by using an AFM tip, for instance. Conventional SQUIDs are used to measure magnetic fields of order of nano-Tesla routinely. A GSQUID of a similar size can detect a geometric curvature with radius $\sim1$ mm which is very large compared to the London length.

Note that the GSQUID is sensitive to the enclosed curvature even though its underlying Cooper pairs only sample its edges.  As a simple example, consider a narrow ribbon around the equator a sphere.  It has zero curvature but encloses a cap with total curvature
$2\pi$ which for $l=1$, i.e., $p$-wave pairing, gives a minus sign in \eqref{mechSQ}. The analogy with the Aharonov-Bohm effect should be obvious.


\textit{Geometric radiator}: An acoustically driven membrane of a chiral superconductor generates, through oscillations of its curvature, an oscillating magnetic dipole moment, and thus it will emit electromagnetic radiation. 

To estimate this effect we approximate the lowest eigenmode of a clamped membrane with a Gaussian profile $h(t,r)=h_0(t)e^{-r^2/r_0^2}$.
For small values of the aspect ratio $ h_0/r_0$, the Gaussian curvature of the bump is $K = (h_0/r_0^2)^2 e^{-r^2/r_0^2}$. 
Based on the Eq. \eqref{glcurrent} we can calculate the current density induced from geometric curvature as
\begin{equation}
\oint J_i dl^i =   \frac{q\gamma}{\hbar} \int d S  K \, ,
\end{equation}
which amounts to an effective magnetic dipole moment,
 \be{magmom}
 m_0 \sim 0.4  \frac{q\gamma d_0 h_0^2}{\hbar}.
 \ee
A magnetic dipole oscillating at a frequency $\Omega$ generates the radiation power\cite{jackson} 
\be{radpow}
\langle P\rangle=\frac{\mu_0 m_0^2 \Omega^4}{12\pi c^3}.
\ee
Taking  $h_0=1 $ mm, $d_0=1 \ \mu$m, and $\rho=10^{-22}$ m$^{-3}$ we get $\langle P\rangle \sim 10^{-6} \left({\Omega} /{\rm THz}\right)^4$ W.
For a thin film, the flexural mode has the lowest acoustic frequency (for graphene this is about 40 THz\cite{taleb}) and may be practical to excite, yielding detectable radiation.

\textit{Summary}:  
We have explored the physical implications of geometric curvature for chiral superconductivity, and found several novel phenomena.  Exploiting the geometric gauge field, we proposed dc and ac geometric Josephson effects, a chiral magnetostriction effect and geometry-based SQUIDs and radiators.  These effects provide both striking signatures for chiral superconductivity and potentially useful devices.

\textit{Acknowledgement}: We gratefully acknowledge helpful discussions with E. Babaev, A. Balatsky, A. Bouhon, T. Kvorning, and J. Carlstr\"{o}m. This work was supported by the Swedish Research Council under Contracts No. 335-2014-7424 and No. 2015-04416.  In addition, F. W.'s work is supported by the U.S. Department of Energy under Grant Contract  No. DE-SC0012567 and by the European Research Council under Grant No. 742104.

\hspace{3mm}


\end{document}